\documentclass[letterpaper, 10 pt, conference]{ieeeconf}  % Comment this line out if you need a4paper

\IEEEoverridecommandlockouts                              % This command is only needed if 
\usepackage{multirow}

\usepackage{amsthm}
\usepackage{amsmath, comment} % assumes amsmath package installed
\usepackage{amssymb} 
\usepackage{graphicx}
\usepackage{algorithm} 
\usepackage[noend]{algpseudocode}

\usepackage{xcolor}

\theoremstyle{definition}

\theoremstyle{remark}

\usepackage{caption}
\usepackage{subcaption}
\usepackage{url}
\usepackage{stackengine,cite}

\usepackage{cancel}
\usepackage{booktabs}
\usepackage{makecell}
\usepackage{placeins}

\title{\LARGE\bf Neural Co-state Regulator: A Data-Driven Paradigm for Real-time Optimal Control with Input Constraints}

\author{Lihan Lian$^{1}$, Yuxin Tong$^{2}$, Uduak Inyang-Udoh$^{2}$ \\
\thanks{$^{1}$Department of Robotics, University of Michigan, Ann Arbor, Michigan, USA. \tt\small lihanl@umich.edu}%
\thanks{$^{2}$Department of Mechanical Engineering, University of Michigan, Ann Arbor, Michigan, USA. \tt\small yuxinton@umich.edu, \tt\small udiinyang@umich.edu}%
}

\begin{document}
\maketitle
\thispagestyle{empty}
\pagestyle{empty}

%%%%%%%%%%%%%%%%%%%%%%%%%%%%%%%%%%%%%%%%%%%%%%%%%%%%%%%%%%%%%%%%%%%%%%%%%%%%%%%%
\begin{abstract}
We propose a novel unsupervised learning framework for solving nonlinear optimal control problems (OCPs) with input constraints in real-time. In this framework, a neural network (NN) learns to predict the optimal co-state trajectory that minimizes the control Hamiltonian for a given system, at any system's state, based on the Pontryagin’s Minimum Principle (PMP). Specifically, the NN is trained to find the norm-optimal co-state solution that simultaneously satisfies the nonlinear system dynamics and minimizes a quadratic regulation cost. The control input is then extracted from the predicted optimal co-state trajectory by solving a quadratic program (QP) to satisfy input constraints and optimality conditions. We coin the term \textit{neural co-state regulator} (NCR) to describe the combination of the co-state NN and control input QP solver. To demonstrate the effectiveness of the NCR, we compare its feedback control performance with that of an expert nonlinear model predictive control (MPC) solver on a unicycle model. Because the NCR's training does not rely on expert nonlinear control solvers which are often suboptimal, the NCR is able to produce solutions that outperform the nonlinear MPC solver in terms of convergence error and input trajectory smoothness even for system conditions that are outside its original training domain. At the same time, the NCR offers two orders of magnitude less computational time than the nonlinear MPC.  %techniques,  it is not subject to We coin the term  As such the Specifically, the neural network is trained to produce co-state trajectory such that the In training, the we propose the design of loss function to train the CoNN more efficiently and coin this paradigm as the neural co-state regulator (NCR). Instead of first solving the optimal co-state trajectories and then training the NN in a supervised learning fashion, CoNN is trained by a better design of loss function based on Pontryagin's Minimum (Maximum) Principle (PMP). Similarly to the work in CoNN, we only extract the first element from the predicted optimal co-state trajectories and subsequently solve a quadratic program (QP) to satisfy both input constraints and optimality. The unicycle model is used as the system dynamics for validation, and comparisons are also made with Model Predictive Control (MPC). The results have shown that the proposed method can greatly improve the speed of computation and also perform well even for unseen initial conditions that are outside the training set and nonzero reference. 

\end{abstract}
% \footnotetext{Supplementary material is available at \url{https://lihanlian.github.io/neural_co-state_regulator/}}

\makeatletter
\renewcommand{\@makefnmark}{}
\makeatother
\footnotetext{Supplementary material is available at \url{https://lihanlian.github.io/neural_co-state_regulator/}}

%%%%%%%%%%%%%%%%%%%%%%%%%%%%%%%%%%%%%%%%%%%%%%%%%%%%%%%%%%%%%%%%%%%%%%%%%%%%%%%%

\section{Introduction} \label{sec-introduction}
\subsection{Motivation and Related Work}

Optimal control has been widely applied in safety-critical domains such as aerospace, robotics, and energy systems \cite{teo2021applied}. Efficiently solving nonlinear optimal control problems (OCPs) is crucial but remains challenging due to the lack of analytic solutions and high computational complexity \cite{peaucelle2010complexity}.

Widely-used approaches for solving OCPs can be broadly categorized into direct and indirect methods \cite{Betts2010}. Direct methods are typically implemented using model predictive control (MPC), which enable real-time feedback control by solving an optimization problem at each time step \cite{Grune2011}. However, nonlinear MPC may suffer from slow convergence of the nonlinear program solver, which compromises \textit{real-time} feasibility \cite{schwenzer2021mpc}. Most implementations rely on direct transcription methods, where the continuous-time OCP is discretized and solved via numerical optimization. Explicit MPC reduces online computation, but its offline computation becomes impractical for high-dimensional systems, and the precomputed policy may not generalize well to unseen initial conditions \cite{Nambisan2024Optimal}.

\vspace{0.5\baselineskip}

In contrast, indirect methods such as pontryagin’s minimum
principle (PMP) provide the necessary optimality conditions by formulating the Hamiltonian involving both state and co-state dynamics, resulting in a two-point boundary value problem (TPBVP) \cite{kirk2004pontryagin}. While theoretically elegant, PMP-based methods are not suitable for feedback control and are highly sensitive to initialization, especially in the presence of state and control constraints or nonconvexities \cite{rao2009survey}. Although penalty-based formulations have been introduced to enforce constraints \cite{pagone2022penalty, bonalli2022sequential}, they often still require solving the TPBVP online, limiting scalability \cite{pereira2021aggregated, feng2024optimal}.

\vspace{0.5\baselineskip}

To overcome these limitations, learning-based approaches
have gained traction for approximating optimal control policies
in a data-driven manner. Prominent among these approaches is reinforcement learning (RL), which is based on dynamic programming principles and aims to derive optimal policies through reward-driven interaction with the environment \cite{sutton1998reinforcement}. One of the most important advantages of RL is that it can be used in a model-free manner, and several algorithms, including Deep Deterministic Policy Gradient (DDPG) and Proximal Policy Optimization (PPO), have been widely used in many challenging control tasks \cite{ddpg, ppo}. However, it remains difficult for a trained RL agent to
transfer their experience to new environments, and these algorithms still struggle to generalize between tasks \cite{cobbe2019quantifying}. 

\vspace{0.5\baselineskip}

%The combination of RL with MPC has also been used to solve discrete time OCP \cite{rl_mpc_tnnls}. However, it essentialy uses NN to learn a better terminal cost function and receding optimization problem still needs to be solved at each time step.

% However, standard RL approaches are often sample-inefficient, difficult to interpret, and struggle to enforce hard constraints, particularly in safety-critical continuous-time systems \cite{Garcia2015SafeRL，cheng2019end}. To address these issues, recent studies have explored the integration of Pontryagin’s Maximum Principle (PMP) into learning-based frameworks. Some approaches incorporate PMP-derived structures as regularization terms or supervisory signals within reinforcement learning (RL) pipelines \cite{eberhard2024pontryagin}, while others embed the Hamiltonian dynamics into policy optimization using neural networks to improve sample efficiency and policy smoothness \cite{gu2022pontryagin， Schiassi2024}. PMP has also been integrated into model based control methods such as MPC and Differential Dynamic Programming (DDP) to improve stability and interpretability \cite{Nambisan2024Optimal, sidhoum2025pontryagin}, this efforts typically rely on accurate models and require online optimization at each time step, which can be computationally intensive for real-time constrained problem. 

Imitation learning (IL) offers an alternative approach by training a neural network (NN) to replicate control policies from pairs of state and input data \cite{underactuated}. The goal is to enable the generation of suitable control input in real-time at a fraction of the computational cost required by an expert OCP solver \cite{chen2020imitation, bojarski2016endtoend}.  A common method in IL is behavior cloning (BC), which directly maps states to actions. However, BC is known to suffer from compounding errors due to distributional drift \cite{ross2011reduction}. Recent work such as ALOHA \cite{zhao2023finegrained} addressed this issue through Action Chunking with Transformers (ACT), and has been successfully applied to tasks such as fine-grained bimanual manipulation on low-cost hardware. Despite these advances, IL continues to face challenges in generalizing to unseen states. This limitation stems from its lack of access to system dynamics and, thus, inability to preserve the structure and interpretability of classical control solutions unless it is augmented with domain-aware regularization \cite{effati2013optimal}.

% Distinct from reinforcement learning approaches, another line of work focuses on embedding the structural properties of PMP into end-to-end differentiable models. Early attempts used trial functions to minimize the residuals of PMP conditions \cite{effati2013optimal}, while more recent efforts such as Pontryagin Differentiable Programming (PDP) embed Hamiltonian dynamics into end-to-end differentiable architectures \cite{pdp_nips_2020}. However, these approaches still require solving TPBVPs either during training or inference, limiting their real-time applicability and scalability to high-dimensional systems.

% A third line of research attempts to directly approximate the TPBVP arising from PMP using NNs \cite{pontryagin_nn-mathematics, zang2022machine}. However, these efforts have been mostly restricted to learning solutions for specific boundary (initial) conditions. In a previous work, the authors have extended this paradigm to more generally parameterize the TPBVP solution for different initial conditions by using the so-called co-state NNs (CoNNs) \cite{lian2025co}. In this parameterization, the CoNN learns the mapping from any given state to its corresponding optimal co-state trajectory using solutions provided by an expert TPBVP solver for training. Nevertheless, in nonlinear problems, expert TPBVP solutions are generally suboptimal and nonunique, especially for multiinput systems, making this supervised learning approach restrictive. 

\vspace{0.5\baselineskip}

A third line of research attempts to directly approximate the TPBVP arising from PMP using NNs \cite{pontryagin_nn-mathematics, zang2022machine}. These efforts typically target specific boundary conditions and focus on learning solutions from them. To generalize beyond fixed initial conditions, a previous work by the authors introduced the concept of co-state neural networks (CoNNs), which learn a mapping from a given state $\mathbf{z}(t)\in \mathbb{R}$ to its corresponding optimal co-state trajectory using supervision from expert TPBVP solvers \cite{lian2025co}. Nevertheless, in nonlinear problems, expert TPBVP solutions are generally suboptimal and nonunique, especially for multi-input systems, making this supervised learning approach restrictive.

\subsection{Contribution}
 %In a previous work, the authors proposed a co-state NN which uses a NN to learn the mapping from a state to its corresponding optimal co-state trajectory in a supervised learning manner \cite{lian2025co}. This requires high quality of training data (state and co-state trajectory pairs), which may be difficult to obtain for higher-dimensional systems, and only one-dimensional nonlinear system is used as an example. 
 In this work, we present a learning framework in which the CoNN is trained in an unsupervised fashion eliminating the need for expert optimal co-state trajectories. This is achieved by use of a loss function that incorporates both the stage and the terminal costs in the corresponding OCP as well as an additional co-state regularization. This loss function's design constrains the NN to find the optimal mapping from a state to the minimum-norm co-state trajectory in a manner that satisfies the system dynamics. The optimal control solution may be subsequently obtained by solving a QP based on the CoNN's predicted optimal co-state trajectory and existing input constraints. In this paper, we focus on an OCP with a quadratic regulation cost; thus, the control learning framework is termed \textit{neural co-state regulator} (NCR). The performance of the NCR is evaluated in a feedback control loop on various scenarios within and outside of the training domain. A comparison with nonlinear MPC is made in all cases. To enumerate, the contributions of this paper is three-fold:
%The NCR is also able to reach non-zero reference for unseen initial conditions, and MPC is conducted for comparison in all cases. 
% Unlike the previous work of CoNN, which is trained in a supervised learning fashion 
% \cite{lian2025co}, the proposed neural co-state regulator (NCR) in this work eliminates the need for explicit co-state computation. Inspired by MPC, our loss function incorporates both stage and terminal cost from the original OCP, along with a regularization term that promotes co-state consistency. This formula encourages the NN to find the optimal mapping from a state to the co-state trajectory while also simplifying the process of generating training dataset. We evaluated NCR on various scenarios, including both seen and unseen initial conditions with input constraints. The NCR is also able to reach non-zero reference for unseen initial conditions, and MPC is conducted for comparison in all cases. The key contributions of the paper are three-fold:

\begin{enumerate}
    % \item We propose a paradigm that trains the CoNN using a properly designed loss function. This not only simplifies the training data set generation, but is also crucial for a system that may be difficult to obtain ground-truth optimal co-state trajectories.
    \item We propose a novel training paradigm for CoNNs using a PMP-informed loss function that eliminates the need for ground-truth optimal co-state trajectories, enabling application to systems where such trajectories are difficult or expensive to obtain.

% \begin{itemize}
    % \item The performance of NCR is validated on a higher dimensional nonlinear system with real engineering application (unicycle model). Our NCR is on par with or outperforms the MPC approach in some cases, with a much faster computational speed.
    \item We validate the proposed NCR on a nonlinear unicycle model with control input constraints and compare its performance to that of MPC. We demonstrate the ability of the NCR to achieve comparable or superior control performance to MPC with significantly reduced computational time.
% \end{itemize}

% \begin{itemize}
    % \item The trained CoNN demonstrates good generalizability in the sense of both unseen initial states and non-zero reference. This makes it able to be used for an NN-based feedback controller for a wider range of situations.
    \item We verify the trained NCR's ability to provide control solution both unseen initial states and nonzero reference, demonstrating its potential as a scalable and efficient neural feedback controller for constrained, nonlinear systems. 
    
% \end{itemize}

\end{enumerate}
\subsection{Paper Structure}
The structure of this paper is organized as follows. Section II provides an overview of the direct and indirect methods for solving OCP in a model-based paradigm. Section III formulates the general finite-horizon OCP that the proposed NCR is designed to address. In Section IV, we elaborate on the neural network architecture, training methodology, and describe the strategy used to enforce control input constraints, along with the procedure to validate the NCR. Section V presents the experimental results for a unicycle model, including a detailed performance comparison between the NCR and the MPC. Finally, Section VI summarizes the key contributions of this work and discusses potential avenues for future research.

% \cite{explicit-mpc-upenn}
% \cite{explicit-mpc-survey}
% \cite{explicit-mpc-nn}
% \cite{pinn-jcp}
%\cite{neural-ode-icl}
%\cite{sdre-acc}
% \cite{sdre-lcss}
%\cite{neural-ode-nips}
% \cite{sdre-ifac}
% \cite{pmp_based_dl-jmlr}
% \cite{pmp_based_dl-pmlr}
%\cite{pontryagin_nn-mathematics}
%\cite{co-state-nn-gatech}
\section{Background} \label{sec-background}
Consider a continuous time OCP with the goal of minimizing a cost functional $\mathcal{J}$ formulated as:
\begin{subequations} \label{eq:background_ocp_formulation} % Label for the whole system
\begin{align}
\mathcal{J} & = \phi(\mathbf{z}(t_f)) + \int_{0}^{t_f} L(\mathbf{z}(t), \mathbf{u}(t), t) \, dt, \\
\text{s.t.} \quad  & \dot{\mathbf{z}}(t) = f(\mathbf{z}(t), \mathbf{u}(t), t),  \label{eq:background_ocp_formulation-state_dynamics} \\
& \mathbf{z}(0) = \mathbf{z}_0, \label{eq:background_ocp_formulation-initial-condition}\\
& \mathbf{u}(t) \in \mathcal{U} \label{eq:background_ocp_formulation-input-constraint},
\end{align}
\end{subequations}
Here, $\mathbf{z}(t) \in \mathbb{R}^p$ and $\mathbf{u}(t) \in \mathbb{R}^q$ denote the system state (or tracking error) and control input, respectively, with $p$ and $q$ being their dimensions. The total cost $\mathcal{J}$ consists of a terminal cost \(\phi(\mathbf{z}(t_f))\) and a stage cost \(L(\mathbf{z}(t), \mathbf{u}(t), t)\), accumulated over the time horizon $[0,t_f]$. A feasible solution must satisfy the dynamics (\ref{eq:background_ocp_formulation-state_dynamics}), initial condition (\ref{eq:background_ocp_formulation-initial-condition}), and the input constraint (\ref{eq:background_ocp_formulation-input-constraint}).

\subsection{Direct Method Solving OCP}
Direct methods solve the continuous time OCP by transcribing the problem into a finite-dimensional optimization through time discretization. The system dynamics Eq.~\eqref{eq:background_ocp_formulation-state_dynamics} are discretized as:
\begin{equation}
    \mathbf{z}_{t+1} = f(\mathbf{z}_t, \mathbf{u}_t),
\end{equation}
where $\mathbf{z}_t \in \mathbb{R}^p$ represents the state vector at time step $t \in \mathbb{Z}^+$, $\mathbf{u}_t \in \mathcal{U} \subset \mathbb{R}^q$ is the admissible control input at time step $t$. This is used to transcribe the OCP Eq.~\eqref{eq:background_ocp_formulation} into an optimization problem with a finite number of decision variables, thus following the scheme of \textit{discretize then optimize} paradigm of solving OCPs \cite{nonlinear-programming-book}.

Direct methods are typically implemented using MPC, which solves an OCP over a finite horizon $P$ at each time step: 
\begin{subequations} \label{eq:mpc_formulation} % Label for the whole system
\begin{align}
J_t^*(\mathbf{z}_t) &= \min_{\mathbf{u}_{t:t+P-1|t}} \ell_f(\mathbf{z}_{t+P|t}) + \sum_{k=0}^{P-1} \ell(\mathbf{z}_{t+k|t}, \mathbf{u}_{t+k|t})  \\
\text{s.t.} \quad  & \mathbf{z}_{t+k+1|t} = f(\mathbf{z}_{t+k|t}, \mathbf{u}_{t+k|t}), \quad k = 0, \dots, P-1  \\
& \mathbf{u}_{t+k|t} \in \mathcal{U}, \quad k = 0, \dots, P-1 \\
& \mathbf{z}_{t|t} = \mathbf{z}_t,  \\
& \mathbf{z}_{t+P|t} \in \mathcal{Z}_f.
\end{align}
\end{subequations}

% The goal is to optimize $J_t^*(\mathbf{z}_t)$ and the input sequence $\mathbf{u}_{t:t+P-1|t}$ are the decision variables. $\mathbf{z}_{t+k|t}$ denotes the state vector at time step $t + k$ predicted at time step $t$, obtained starting from the current state $\mathbf{z}_t$. The terminal cost and stage cost are represented as $\ell_f(\mathbf{z}_{t+P|t})$ and $\ell(\mathbf{z}_{t+k|t}, \mathbf{u}_{t+k|t})$, respectively, $\mathcal{Z}_f$ is the terminal set and $P$ is the receding horizon. This process is repeated at each time step and the size of this optimization problem grows as the horizon $P$ increases. 

The goal is to minimize the cost function $J_t^*(\mathbf{z}_t)$, where the control inputs $\mathbf{u}_{t:t+P-1|t}$ serve as the decision variables over the receding horizon $P$. At each time step $t$, the state vector $\mathbf{z}_{t+k|t}$ is predicted forward from the current state $\mathbf{z}_t$ based on the system dynamics. The cost is composed of two parts: a terminal cost $\ell_f(\mathbf{z}_{t+P|t})$ at the end of the horizon, and a stage cost $\ell(\mathbf{z}_{t+k|t}, \mathbf{u}_{t+k|t})$ accumulated over $k = 0, \dots, P-1$. The terminal state $\mathbf{z}_{t+P|t}$ is constrained to lie within a terminal set $\mathcal{Z}_f$. This optimization is repeated at every time step, and the problem size increases with the horizon length $P$.

\subsection{Indirect Method Solving OCP}
Indirect methods solve optimal control problems by first deriving the necessary conditions, thereby preserving the problem's structure before discretization. This aligns with the paradigm of \textit{ optimize and then discretize} \cite{nonlinear-programming-book}. 

A central result underpinning this class of methods is PMP, which provides first-order necessary conditions for systems governed by differential equations. For the OCP in Eq.~\eqref{eq:background_ocp_formulation}, PMP introduces the control Hamiltonian $H$, defined as:
\begin{align}
H(\mathbf{z}(t), \mathbf{u}(t), \mathbf{\lambda}(t), t) &= L(\mathbf{z}(t), \mathbf{u}(t), t)\notag \\ 
 & + \mathbf{\lambda}^\top(t) f(\mathbf{z}(t), \mathbf{u}(t), t),
\label{eq:hamiltonian}
\end{align}
where \(\mathbf{\lambda}(t) \in \mathbb{R}^n\) is the co-state (or adjoint) vector. The state and co-state dynamics are derived by the following differential equations:
\begin{equation}
\dot{\mathbf{z}}(t) = \nabla_{\mathbf{\lambda}} H,
\label{eq:state_dynamics}
\end{equation}
\begin{equation}
\dot{\mathbf{\lambda}}(t) =  -\nabla_{\mathbf{z}} H.
\label{eq:costate_dynamics}
\end{equation}
Since the OCP \eqref{eq:background_ocp_formulation} has a fixed final time and a free final state, the terminal boundary condition for co-state variable can be obtained as follows: 

\begin{equation*}
 \mathbf{\lambda}(t_f) = \nabla_\mathbf{z} \phi(\mathbf{z}(t_f)).
\label{eq:co-state_terminal_condition}
\end{equation*}

Given sufficient boundary conditions involve both state and co-state, the corresponding TPBVPs can then be solved. In the case of constrained control input, PMP states that the optimal control \(\mathbf{u}^*(t)\) minimize the Hamiltonian $H$:
\begin{equation}
\mathbf{u}^*(t) = \arg\min_{\mathbf{u}(t) \in \mathcal{U}} H(\mathbf{z}^*(t), \mathbf{u}(t), \lambda^*(t), t).
\label{eq:optimal_control}
\end{equation}

Even without input constraints, solving the TPBVPs can be challenging as analytical solutions rarely exist. Numerical techniques such as shooting methods are typically used \cite{numerical-tpbvp-book}, \cite{collocation-method-ocp}.%, and an additional challenge can be imposed by the higher-dimensional system state.

\section{Problem Statement} \label{sec-problem-statement}

Consider a control-affine system governed by a finite-horizon optimal control problem with quadratic stage cost in continuous time. The objective is to minimize the cost functional: 
\begin{subequations} \label{eq:ocps_formulation} % Label for the whole system
\begin{align}
\min \quad J & = \int_{0}^{t_f} \left( \mathbf{z^\top}Q\mathbf{z} + \mathbf{u^\top}R\mathbf{u} \right) dt + \phi(\mathbf{z}(t_f)), \\
\text{s.t.} \quad & \mathbf{\dot z}(t) = f(\mathbf{z}(t)) + g(\mathbf{z}(t))\mathbf{u}(t), \\
& \mathbf{z}(0) \in \mathcal{Z}\\ 
& \mathbf{u}(t) \in \mathcal{U}.  
\end{align}
\end{subequations}

% Here, $\phi(\mathbf{z}(t_f))$ defines the quadratic terminal cost, while $\mathcal{U}$ denotes the set of admissible control inputs. The vectors $\mathbf{z(0)}$ and $\mathbf{z}(t_f)$ represent the initial and terminal states, respectively, where $\mathcal{Z}$ is a set of possible initial conditions \footnote{Note that the problem formulation here differs from standard OCPs, where $\mathbf{z}(0)$ is fixed. This is to emphasize that the OCP admits a family of solutions.}. $f(\mathbf{z})$ and $g(\mathbf{z})$ are functions with appropriate dimensions that describe the dynamics of the system. The quadratic stage cost is characterized by the weighting matrices \(Q \in \mathbb{R}^{p \times p}\) and \(R \in \mathbb{R}^{q \times q}\), where $Q$ is a semidefinite symmetric matrix and $R$ is a symmetric positive definite matrix. 

Here, $\phi(\mathbf{z}(t_f))$ defines the quadratic terminal cost. The admissible control inputs belong to the set $\mathcal{U}$, while $\mathbf{z}(0)$ and $\mathbf{z}(t_f)$ represent the initial and terminal states, respectively. $\mathcal{Z}$ is a set of possible initial conditions \footnote{Note that the problem formulation here differs from standard OCPs, where $\mathbf{z}(0)$ is fixed. This is to emphasize that the OCP admits a family of solutions.}. The system dynamics are described by the functions $f(\mathbf{z}(t))$ and $g(\mathbf{z}(t))$, which have appropriate dimensions. The quadratic stage cost is characterized by the weighting matrices \(Q \in \mathbb{R}^{p \times p}\) and \(R \in \mathbb{R}^{q \times q}\), where $Q$ is a semi-definite symmetric matrix and $R$ is a symmetric positive definite matrix.

Following PMP, the Hamiltonian associated with this optimal control problem is given by: 

\begin{align}
H(\mathbf{z}(t), \mathbf{u}(t), \mathbf{\lambda}(t), t) &= \mathbf{z}^\top(t)Q\mathbf{z}(t) + \mathbf{u}^\top(t)R\mathbf{u}(t) + \notag \\
& \mathbf{\lambda}^\top(t) \left( f(\mathbf{z}(t)) + g(\mathbf{z}(t))\mathbf{u}(t)\right),
\end{align}
where $\mathbf{\lambda}(t)\in \mathbb{R}^p$ is the co-state vector, which has the same dimension as state vector $\mathbf{z}(t)$. Following Eq.~\eqref{eq:state_dynamics} and Eq.~\eqref{eq:costate_dynamics}, the state equation and co-state equation are derived as follows: 
\begin{equation}
    \dot{\mathbf{z}}(t) = \nabla_{\mathbf{\lambda}} H = f(\mathbf{z}(t)) + g(\mathbf{z}(t))\mathbf{u}(t),
\end{equation}
\begin{align}
    \dot{\mathbf{\lambda}}(t) = -\nabla_{\mathbf{z}}{H} &= -2Q\mathbf{z}(t) - \nabla^\top_{\mathbf{z}}f(\mathbf{z}(t))  \mathbf{\lambda}(t) \notag \\
    &\quad -  \nabla^\top_{\mathbf{z}}\left(g(\mathbf{z}(t)) \mathbf{u}(t) \right) \mathbf{\lambda}(t).
\end{align}
After solving the optimal $\mathbf{z}^*(t)$ and $\mathbf{\lambda}^*(t)$ from the corresponding TPBVP, the optimal control input \(\mathbf{u}^*(t) \) is obtained from Eq.~\eqref{eq:optimal_control}. Given the quadratic cost defined by $Q$ and $R$, when there are no constraints on the control input, the optimal control law can be obtained by setting: 
\begin{equation} \label{eq:unconstrained-optimal-u-partial-H-partial-u}
     \nabla_{\mathbf{u^*}}{H} = \mathbf{0} ,
\end{equation}
which yields
\begin{equation} \label{eq:unconstrained-optimal-u-expression}
    \mathbf{u}^*(t) = -\frac{1}{2} R^{-1}g^\top(\mathbf{z}(t)) \mathbf{\lambda}^*(t).
\end{equation}
When there are constraints, the optimal control input can be obtained from:
\begin{equation} \label{eq:constrained-optimal-u}
\mathbf{u}^*(t) = \arg\min_{\mathbf{u}(t) \in \mathcal{U}} \left( \mathbf{u}^\top R\mathbf{u} + \mathbf{\lambda}^{*\top}(t) g(\mathbf{z}) \mathbf{u} \right).
\end{equation}
% \begin{equation} \label{eq:constrained-optimal-u}
% \mathbf{u}^*(t) = \arg\min_{\mathbf{u}(t) \in \mathcal{U}} \left( \mathbf{u}^\top(t)R\mathbf{u}(t) + \mathbf{\lambda}^{*\top}(t) g(\mathbf{z}(t)) \mathbf{u}(t) \right).
% \end{equation}

% One of the major differences between this work and previous work on CoNN is that the proposed paradigm of NCR does not require a ground-truth optimal co-state trajectory as training data. This is indeed beneficial since, in many instances, the optimal co-states cannot be easily obtained. In addition, even if the co-state trajectory is solved by the numerical solver, it may contain a mix of optimal and sub-optimal solutions, which can reduce the training data quality and impose a negative impact on training convergence. Thus, the paradigm proposed in this work simplifies the process of generating training data while still keeping the CoNN capable of seeking a family of solutions $\forall \mathbf{z}(0) \in \mathcal{Z}$.

% \textcolor{red}{In this paper, we propose a paradigm that allows CoNN to parameterize the mapping from a range of state to their corresponding optimal co-state trajectory in a more efficient way. This also strengthens the CoNN to deal with higher-dimensional nonlinear systems that may be challenging to solve for optimal co-states and can be used as feedback controller for a wider range of application. }

To avoid the computational expense associated with numerically solving the TPBVP for real-time control applications, we seek a CoNN (co-state NN) that parametrizes the TPBVP solution for any given initial condition. Moreover, we desire a training framework for the NN which will enables it to learn optimal co-state trajectories without relying on ground-truth supervision. Such a framework will bypass the need for solving the boundary value problem during training and mitigates training issues caused by sub-optimal numerical solutions, and yet allows us to parameterize the solution to the TPBVP $\forall \mathbf{z}(0) \in \mathcal{Z}$. In the following section, we propose a NCR framework that will enable us achieve this objective and allow for efficient feedback control based on the PMP. 

%learning across a family of initial states $\forall \mathbf{z}(0) \in \mathcal{Z}$, even for higher-dimensional nonlinear systems. 

\section{Methodology}\label{sec-methodology}

The CoNN used in NCR parametrizes the mapping from a state to its corresponding optimal co-state trajectories. In this section, we detail the CoNN's  architecture and training procedure, and describe how the control input is extracted from the CoNN in a feedback control context.  %process of how to obtain such an NCR and its usage, this section outlines the CoNN's  architecture, training procedure, acquisition of control input, and performance validation.
\begin{figure*}[t]  % Use figure* for spanning two columns
    \centering
    \includegraphics[trim=0cm 0.0cm 0.0cm -0.8cm,width=1\textwidth]{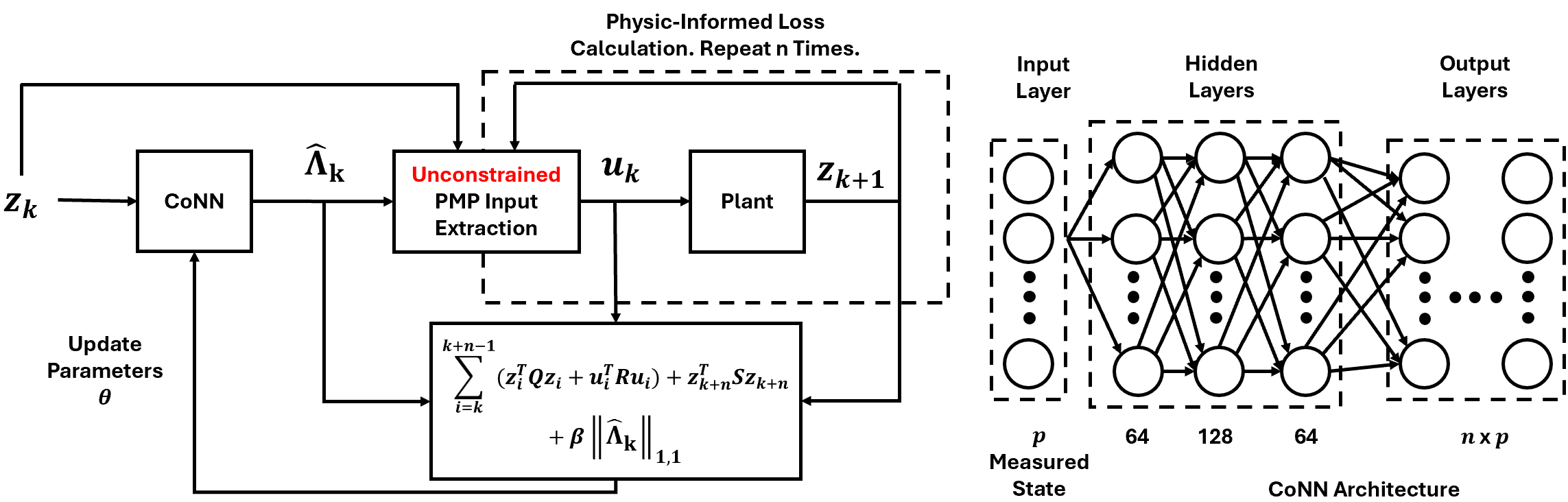}
    \caption{NCR training procedures and CoNN architecture. Given the state $\mathbf{z_k} \in \mathbb{R}^p$, the CoNN predicts optimal co-state values $\boldsymbol{\hat{\Lambda}}_{k} \in \mathbb{R}^{n \times p}$ over a horizon of $n$ time steps. Control input $\mathbf{u_k}$ is evaluated using ($\mathbf{z_k}$, $\boldsymbol{\hat{\lambda}}_{k}$) based on the PMP, applied to the plant to produce the next state $\mathbf{z_{k+1}}$, and the process is repeated. The resulting trajectories are used to compute a loss function to update the parameters of the CoNN. Note that $\mathbf{z_k}$ represents one sample from the training data points and the parameters update illustrated here is for one sample in one training epoch. }
    \label{fig:architecture}
\end{figure*}

\subsection{Neural Network Architecture}
The CoNN is essentially a simple fully connected feedforward neural network, which takes a state vector and predicts a co-state trajectory. At each time step $k$, for a state vector $\mathbf{z_k} \in \mathbb{R}^p$ and a finite horizon of CoNN prediction $n \in \mathbb{Z}^+$, the output of CoNN is a co-state trajectory $\boldsymbol{\hat{\Lambda}}_{k}$ with dimension $n \times p$ as illustrated in Fig. \ref{fig:architecture}.

\begin{equation}
\boldsymbol{\hat{\Lambda}}_{k} = \text{CoNN}_{\theta}(\mathbf{z_k}),
\end{equation}

where $\theta$ denotes the CoNN parameters. $\boldsymbol{\hat{\Lambda}}_{k}$ represents the co-state vectors starting from the current time step $k$ and ending at $k+n-1$, which means $\boldsymbol{\hat{\Lambda}}_{k} = [\boldsymbol{\hat{\lambda}}_{k}, \boldsymbol{\hat{\lambda}}_{k+1}, \dots , \boldsymbol{\hat{\lambda}}_{k+n-1}]$. The co-state vector $\boldsymbol{\hat{\lambda}}_{k} = [\hat{\lambda}_{1,k}, \dots, \hat{\lambda}_{p,k}]^\top \in \mathbb{R}^p$, which is of the same size as the state vector $\mathbf{z_k}$.

\subsection{Training Procedures}

\subsubsection{Training Data Points} 
Given the set of possible state conditions shown in Eq.~\eqref{eq:ocps_formulation}, a reasonable number $N$ of data points are sampled from $\mathcal{Z}$. For the unicycle model used in the following example section, we use $1000$ evenly sampled states to generate the training data points $\boldsymbol{z_{train}}$.

\subsubsection{Loss function} Reinforcement learning is often used to solve sequential decision making problems based on Bellman's principle of optimality and is closely related to optimal control \cite{sutton1998reinforcement, mehta2009qlearning}. Among various RL algorithms, Deep Q-Learning is a popular method that approximates the state-action value function using the Deep Q-Network (DQN) \cite{deep-q-learning}. Without relying on labeled ground truth, DQN improves its approximation by minimizing temporal difference (TD) loss during training. Motivated by this concept, we designed a loss function for CoNN that draws on PMP guidance, enabling the learning process to be aligned with the structure of continuous time optimal control and used in Algorithm \ref{alg:alg1}.

The loss function consists of three components. The first component is stage loss $\mathcal{L_{\text{stage}}}$, which mirrors the stage cost in the context of MPC. Note that the control input constraints are not taken into account during training, since the optimality of the co-state is independent of the input constraints. Consequently, $\mathbf{u_k}$ can be obtained analytically based on Eq.~\eqref{eq:unconstrained-optimal-u-expression}. The stage loss is defined as the cumulative cost over a finite horizon, expressed as:

\begin{equation} \label{eq:L_stage}
\mathcal{L}_{\text{stage}} = \sum_{i=k}^{k+n-1} \left( \mathbf{z_i}^\top Q\mathbf{z_i} + \mathbf{u_i}^\top R\mathbf{u_i} \right).
\end{equation}

Analogously to the terminal cost in MPC formulations, the second component of the loss function is the terminal loss, which penalizes the final state at the end of the horizon. This term $\mathcal{L}_{\text{terminal}} = \phi(\mathbf{z_{k+n}})$ is defined using the same terminal cost function as in the corresponding OCP in \eqref{eq:ocps_formulation}.

The third component of the loss function is the regularized co-state loss \footnote{For an $m\times n$ matrix $A$, we use the entry-wise matrix norm to define the loss as follows: $\|A\|_{p,p} = \|\text{vec}(A)\|_p = \left( \sum_{i=1}^{m} \sum_{j=1}^{n} |a_{ij}|^p \right)^{1/p}$.}, 
which is used to regulate co-state trajectories toward zero at the end of the prediction horizon. It is defined as the sum of the absolute value of all co-state trajectory entries, multiplied by a constant scalar hyperparameter $\beta$,
\begin{equation*} \label{eq:L_co-state}
\mathcal{L}_{\text{co-state}} = \beta \|\boldsymbol{\hat{\Lambda}}_{k}\|_{1,1}.
\end{equation*}
\begin{algorithm}
\caption{NCR Training Procedures}\label{alg:alg1}
\begin{algorithmic}[1] % Adds line numbers
\Procedure{CoNN Training}{}
    \State \textbf{Define} total training epochs $N_{\text{epoch}}$, co-state trajectory prediction horizon $n$, regularized co-state loss hyperparameter $\beta$, learning rate $\alpha$ and initialize CoNN parameters $\theta$
    \State \textbf{For} $e$ in $range(N_{\text{epoch}})$ \textbf{do}
        \State \quad \textbf{For} $\mathbf{z_k} \in \boldsymbol{z_{train}}$ \textbf{do}

            \State \quad \quad  $\boldsymbol{\hat{\Lambda}}_{n} = \text{CoNN}_{\theta}(\mathbf{z_k})$
            \State \quad \quad  Obtain $[\mathbf{u_k}, \dots \mathbf{u_{k+n-1}}]$ based on PMP
            \State \quad \quad Calculate $\mathcal{L}_{\text{stage}}$, $\mathcal{L}_{\text{terminal}}$ and $\mathcal{L}_{\text{co-state}}$
            \State \quad \quad \textbf{Update} CoNN parameters $\theta$
            
        \State \quad \textbf{End for}
    \State \textbf{End for}
\EndProcedure
\end{algorithmic}
\end{algorithm}
\subsection{Control Input Constraints Handling}
Consider that the configuration of OCP described by ~\eqref{eq:ocps_formulation}, when there are no input constraints, the optimal control policy can be obtained based on Eq.~\eqref{eq:unconstrained-optimal-u-expression} provided that CoNN predicts the optimal co-state trajectory. In case of constrained input, the optimal control policy can be obtained by solving a simpler QP illustrated in Eq.~\eqref{eq:constrained-optimal-u}. In our case, we extract the first element of co-state trajectory predicted by CoNN and QP should then be solved as follows:

\begin{subequations} \label{eq:easier_qp} % Label for the whole system
\begin{align}
    u^*_k &= \arg\min \left(\mathbf{u_k^\top} R \mathbf{u_k} + \hat{\boldsymbol{\lambda}}_{k}^\top g(\mathbf{z_k}) \mathbf{u_k} \right), \\
    \text{s.t.} \quad & \boldsymbol{\hat{\Lambda}}_{k} = \text{CoNN}_{\theta}(\mathbf{z_k}),  \\
    & \hat{\boldsymbol{\lambda}}_{k} = \boldsymbol{\hat{\Lambda}}_{k}[0], \\
    & \mathbf{u_k} \in \mathcal{U}.
\end{align}
\end{subequations}
% where $\hat{\boldsymbol{\lambda}}_{\text{traj}, n}^{k}$ is a .
\subsection{NCR Validation}

\begin{figure}[h!]
    \centering    \includegraphics[trim=0cm 0.0cm 0.0cm -0.8cm,width=0.5\textwidth]{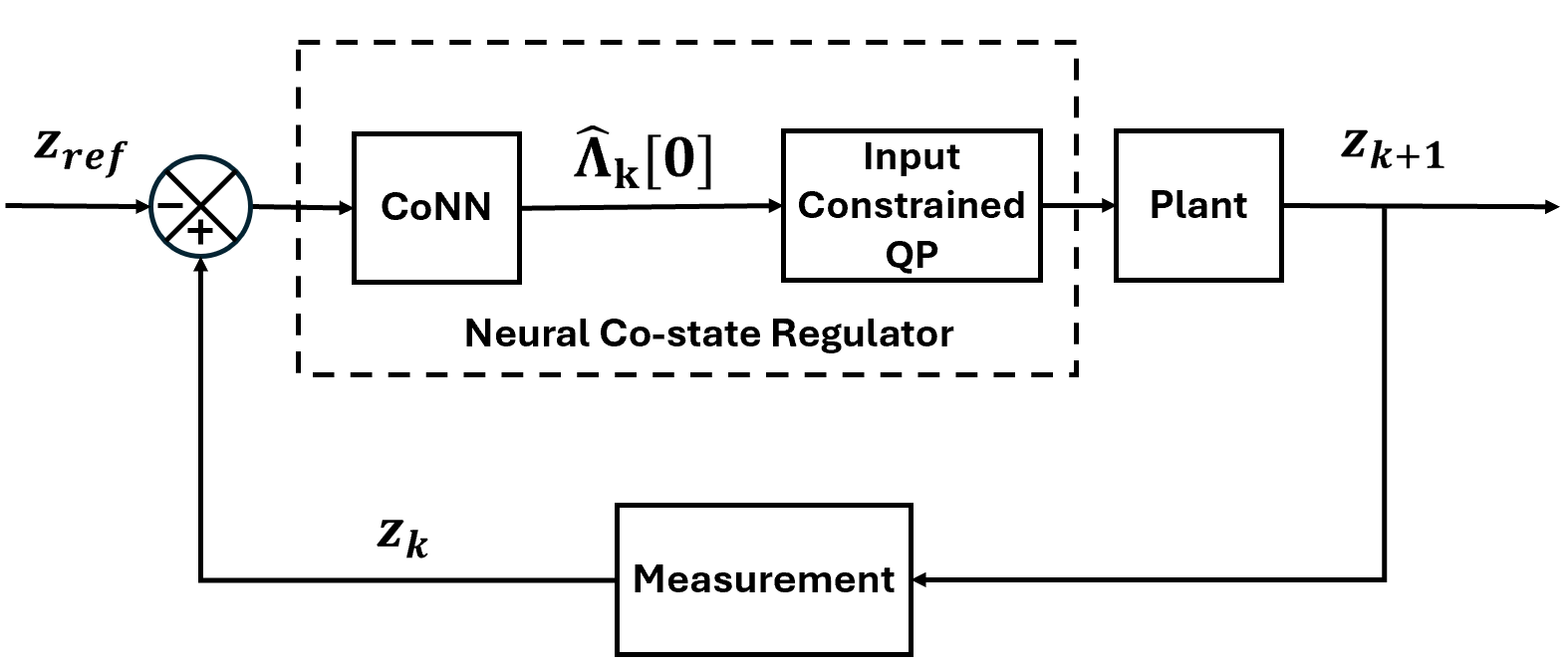}
    \caption{Neural co-state regulator validation block diagram. Only the first element of predicted co-state trajectory is used for solving a QP to obtain constrained optimal input.}
    \label{fig:validation}
\end{figure}

NCR is evaluated in a real-time control loop using simulation as illustrated in Fig. \ref{fig:validation}. At each time step $k$, CoNN takes the state value $\mathbf{Z}_k$ as input and predicts the corresponding optimal co-state trajectories $\boldsymbol{\hat{\Lambda}}_{k}$ with a length of $n$ for each state variable, which is a matrix of dimension $n\times p$. Only the first element of $\boldsymbol{\hat{\Lambda}}_{k}$, a co-state vector $\hat{\boldsymbol{\lambda}}_{k}$ is extracted and used to solve an easier QP based on Eq.~\eqref{eq:easier_qp} to obtain the optimal control input. We use a fourth-order Runge-Kutta integrator to step the system one time step forward and start the simulation at the next time step.

% \textcolor{red}{Similarly, during the training process in this work, the CoNN is trained in a \textit{receding horizon fashion} using the dataset generated from offline optimal solutions of a numerical solver. Thus, the predicted co-state trajectory should resemble the similar optimized properties/pattern??? from MPC that solved online. In realtime operation, only the first element from the predicted co-state trajectory is used for solving control input.} 

 % During validation, the CoNN-based controller operates in discrete time and optimal control input can be obtained by solving a much easier QP with less decision variable. 
\section{Example}\label{sec-example}
Consider the following nonlinear optimal control problem for a unicycle model in continuous time that has quadratic stage cost: 
\begin{subequations} \label{eq:example_problem} % Label for the whole system
\begin{align}
\min_{u} \quad J  & = \int_{0}^{t_f} \left( \mathbf{z}^\top Q\mathbf{z} + \mathbf{u}^\top R\mathbf{u}\right) \, dt + \phi(\mathbf{z}(t_f)),\\
\text{s.t.} \quad &  \dot{z_1} = \dot{x} = v \cos(\theta),\\
&  \dot{z_2} = \dot{y} = v \sin(\theta),\\
&  \dot{z_3} =\dot{\theta} = \omega,\\
& \mathbf{u} \in \mathcal{U}, \\
& \mathbf{z}(0) \in \mathbb{R}^3. 
\end{align}
\end{subequations}

Here we denote $z_1 = x, z_2 = y, z_3 = \theta, u_1 = v, u_2 = w$. The control input constraints are $-1 \leq v \leq 1,  -4 \leq \omega \leq 4,$ the same as the constraints used in \cite{rl_mpc_tnnls}. $t_f$ is the wall-clock length of time for both the MPC and the NCR prediction horizon. For the cost function, $Q = diag(10,10,10), R = diag(1,1)$ and $\phi(\mathbf{z}(t_f)) = \mathbf{z}^\top(t_f)S\mathbf{z}(t_f)$, where $S = 50Q=diag(500,500,500)$. 

In the simulation set-up, we use the sampling time $dt = 0.05 s$ and prediction horizon $n = 30$. Thus, $t_f = 1.5s$ in this case. \textit{CasADi} is used as the optimization solver for MPC and, specifically, \textit{ipopt} is the algorithm for solving NLP. During the CoNN training stage, we choose $\beta = 0.1$ for $\mathcal{L}_{\text{co-state}}$. Training data is uniformly sampled from $[-2, 2]$ for all $x, y$ and $\theta$. A total of 1000 ($10 \times 10 \times10$) states are used for training with 10 samples for each state variable. 

Based on PMP, the control Hamiltonian $H$ for this OCP can then be expressed as follows:
\begin{align*}
H  &= 10x^2 + 10y^2 + 10\theta^2 + v^2 + w^2 \notag \\
 & + \lambda_1 v \cos(\theta) + \lambda_2 v \sin(\theta) + \lambda_3 w.
\end{align*}

The differential equation for each optimal co-state variable is derived as follows:

\begin{equation*}
\dot{\lambda}_1^* = - 20 x^*,
\end{equation*}
\begin{equation*}
\dot{\lambda}_2^* = - 20 y^*,
\end{equation*}
\begin{equation*}
\dot{\lambda}_3^* = - 20 \theta^* + \lambda^*_1 v \sin(\theta) - \lambda^*_2v\cos(\theta).
\end{equation*}
The optimal control law for each control input when there are no constraints can be determined analytically based on Eq.~\eqref{eq:unconstrained-optimal-u-partial-H-partial-u} and Eq.~\eqref{eq:unconstrained-optimal-u-expression},
\begin{equation*}
\frac{\partial H}{\partial v^*} = 2v^* + \lambda_1^*\cos(\theta) + \lambda_2^*\sin(\theta) = 0 , 
\end{equation*}
\begin{equation*}
\frac{\partial H}{\partial w^*} = 2w^* + \lambda_3^* = 0,
\end{equation*}
\begin{equation*}\label{eq:optimal_v}
v^* = -\frac{1}{2}(\lambda_1^* \cos(\theta) + \lambda_2^*\sin(\theta)),
\end{equation*}
\begin{equation*}\label{eq:optimal_w}
w^* = -\frac{1}{2} \lambda_3^*.
\end{equation*}
In the case of control input constraints exist, the optimal control input should be solved by minimizing the Hamiltonian $H$ while still satisfy the constraints in Eq.~\eqref{eq:example_problem} at each time step $k$. For the optimization problem, the state variables are constants and $v_k, w_k$ are the decision variables we seek to find the optimum, thus the terms only involve state variables can be ignored. Hence the optimal control input $\mathbf{u^*_k} ([v^*_k, w^*_k]^\top)$ can be obtained by solving an easier QP below:
\begin{subequations} \label{eq:easier_qp_example} % Label for the whole system
\begin{align*}
    \mathbf{u^*_k} &= \arg\min \left( v_k^2 + w_k^2 + \lambda^*_{1,k} v_k \cos(\theta)\right. \notag \\
    & + \left.  \lambda^*_{2,k} v_k \sin(\theta) + \lambda^*_{3,k} w_k\right), \\
    \text{s.t.} 
    & -1 \leq v_k \leq 1, \\
    & -4 \leq \omega_k \leq 4,
\end{align*}
\end{subequations}

where all $\lambda^*_{1,k}, \lambda^*_{2,k}, \lambda^*_{3,k}$ are extracted from the prediction of CoNN, which was trained following the procedures illustrated in Algorithm \ref{alg:alg1}. NCR is then validated in a real-time feedback control loop, as shown in Fig.~\ref{fig:validation}. This means that CoNN takes the state vector $\mathbf{z_k}$ at every time step as input and outputs a co-state trajectory, but only the first element of the output trajectory is used to solve the optimal control input $\mathbf{u^*_k}$ of that particular time step. 

Our NCR is on par with the performance MPC in terms of convergence error and outperforms the MPC in some cases. In all cases, NCR operates with a much faster speed compared to MPC and also demonstrates good generalizability when it encounters state values that are not within the range of training data and non-zero reference. Both simulation and training ($N_{epoch}=50$) are performed on a computer with an i7 CPU and an RTX 4070 GPU.

\textit{Remark} :  Note that for differentially flat nonlinear systems, including the unicycle model or quadrotor systems, flatness-based MPC (FMPC) could be used to avoid solving a nonlinear program and has the computational advantage. However, in order to obtain the convexity of transformed OCP using FMPC, developing a new cost function may be required, and nonlinear parameterization of convex constraints on the original system input $\mathbf{u_k}$ may not map to the convex constraints on the flat input \cite{fmpc-iros}. Thus, for the sake of comparison in a more general scenario, we used the typical MPC approach as a benchmark.

\subsection{Constrained Control Input with Seen $\mathbf{z}(0)$}

\begin{figure}[h!]
\begin{subfigure}[b]{1\columnwidth}
   \begin{center}
   \includegraphics[trim=0cm 0.0cm 0.0cm 0.0cm,width=1\textwidth]{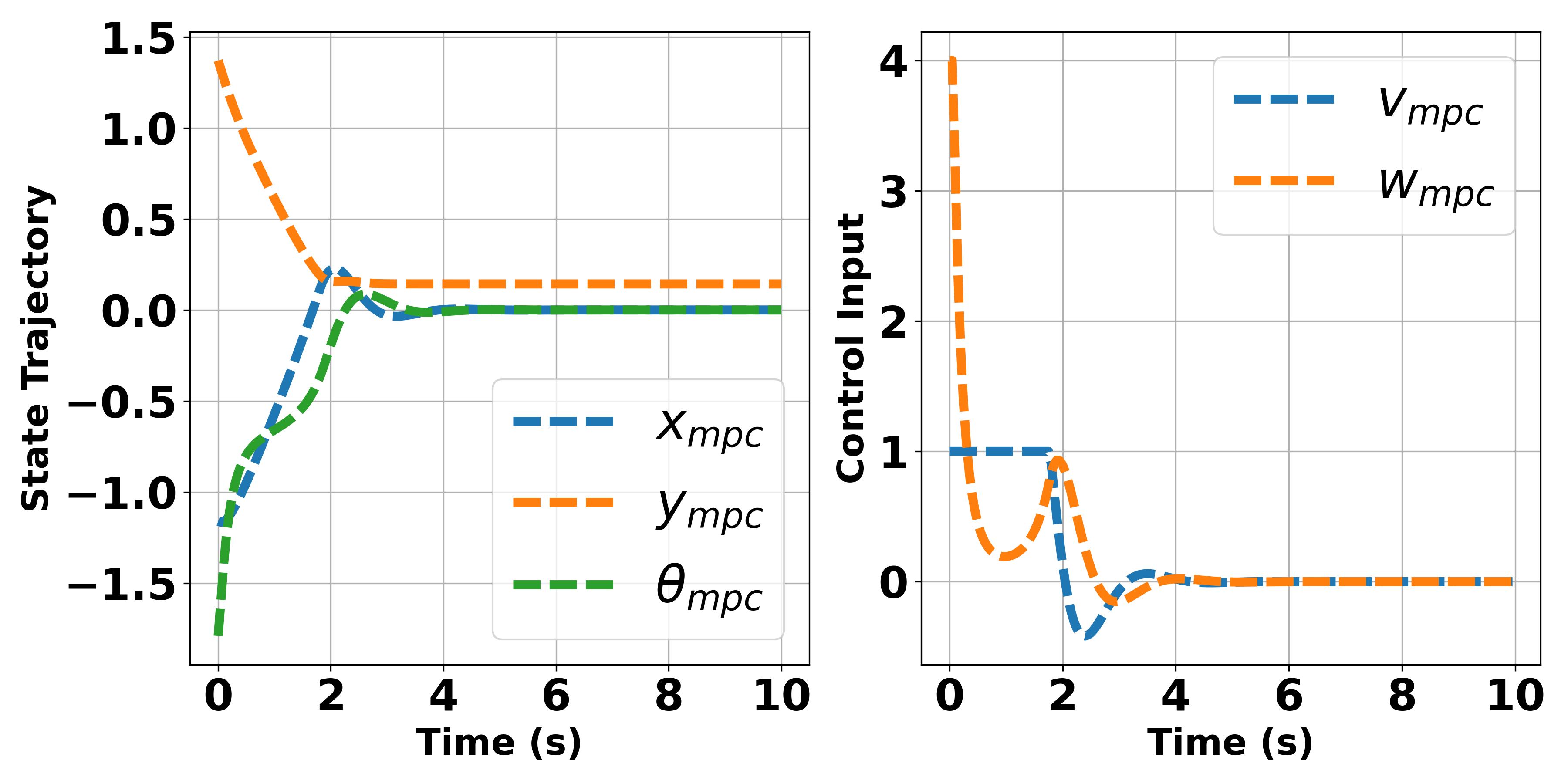}
   \caption{MPC state and control input trajectories.}
   \label{fig:simulation1_mpc}
   \end{center}
\end{subfigure}
\begin{subfigure}[b]{1\columnwidth}
   \begin{center}
      \includegraphics[trim=0cm 0cm 0cm 0cm, width=1\textwidth]{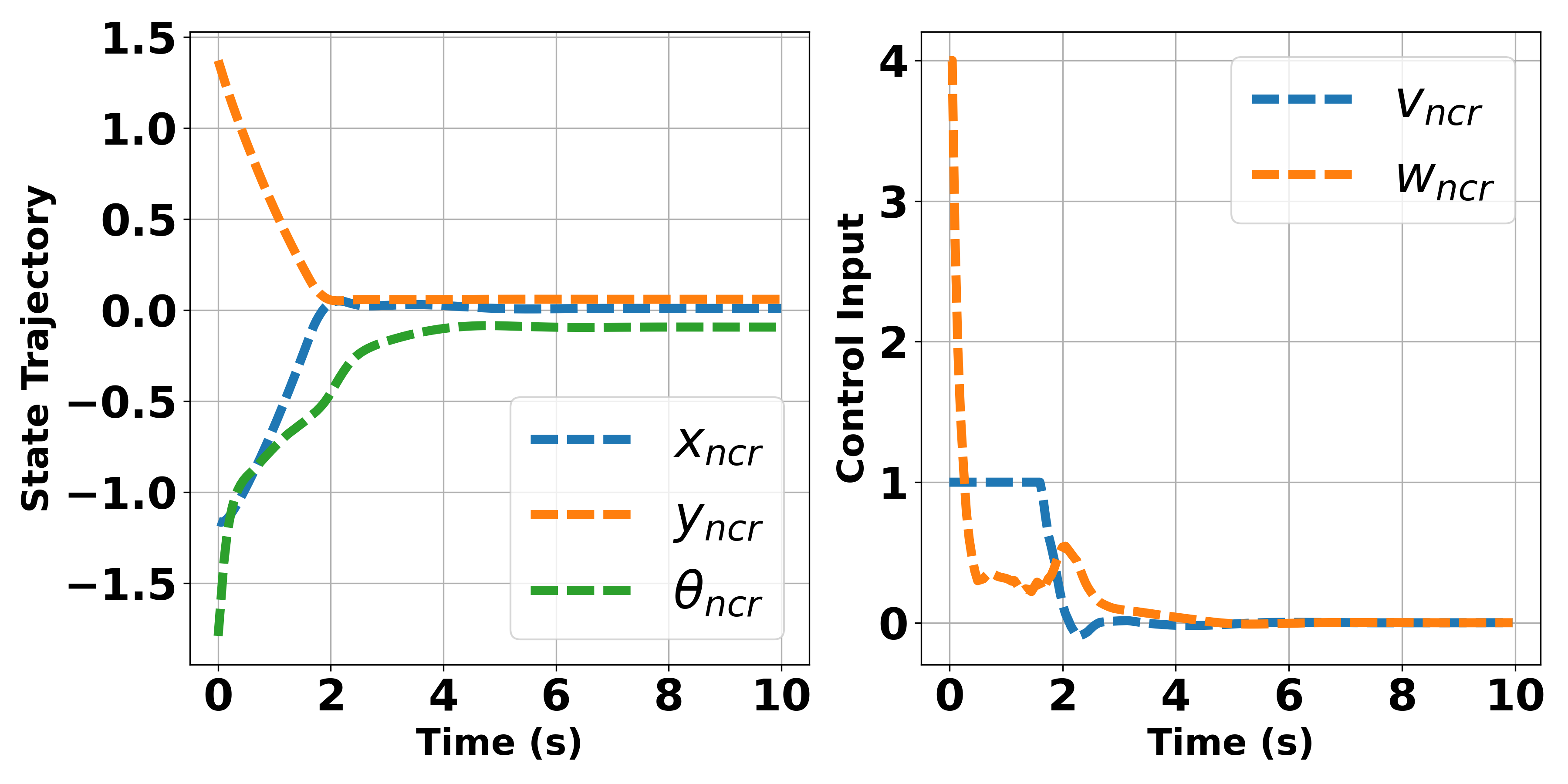}
\caption{NCR state and control input trajectories.}
   \label{fig:simulation1_ncr}
   \end{center}
\end{subfigure}
\caption[8pt]{Comparison of MPC and NCR solutions for initial conditions seen during the NCR training and zero target reference.}\label{fig:simulation1}
\end{figure}

The performance of NCR is first validated in the scenario when the initial condition $\mathbf{z}(0)$ is within the training range. The same prediction horizon $n=30$ is used for both the MPC and CoNN training stage. As shown in Fig. \ref{fig:simulation1}, the same initial condition $\mathbf{z}(0) = [-1.16, 1.37, -1.79]^\top$ is used for both MPC and NCR in close-loop verification. The state trajectory result in $\mathbf{z}(t_f)_{mpc} = [0.00, 0.14, 0.00]^\top$ at the final time for MPC and $\mathbf{z}(t_f)_{ncr} = [0.01, 0.06, -0.09]^\top$ for NCR. NCR is on par with MPC in terms of state convergence error, but with a much faster computational speed at $1.8 ms$ per simulation step compared to $226.6 ms$ for MPC. 
% Both the state and control input trajectories of the NCR-based controller are also smoother (less overshoot) compared to MPC, with quantitative analysis shown in the table below.

\subsection{Constrained Control Input with Unseen $\mathbf{z}(0)$}

\begin{figure}[h!]
\begin{subfigure}[b]{1\columnwidth}
   \begin{center}
   \includegraphics[trim=0cm 0.0cm 0.0cm 0.0cm,width=1\textwidth]{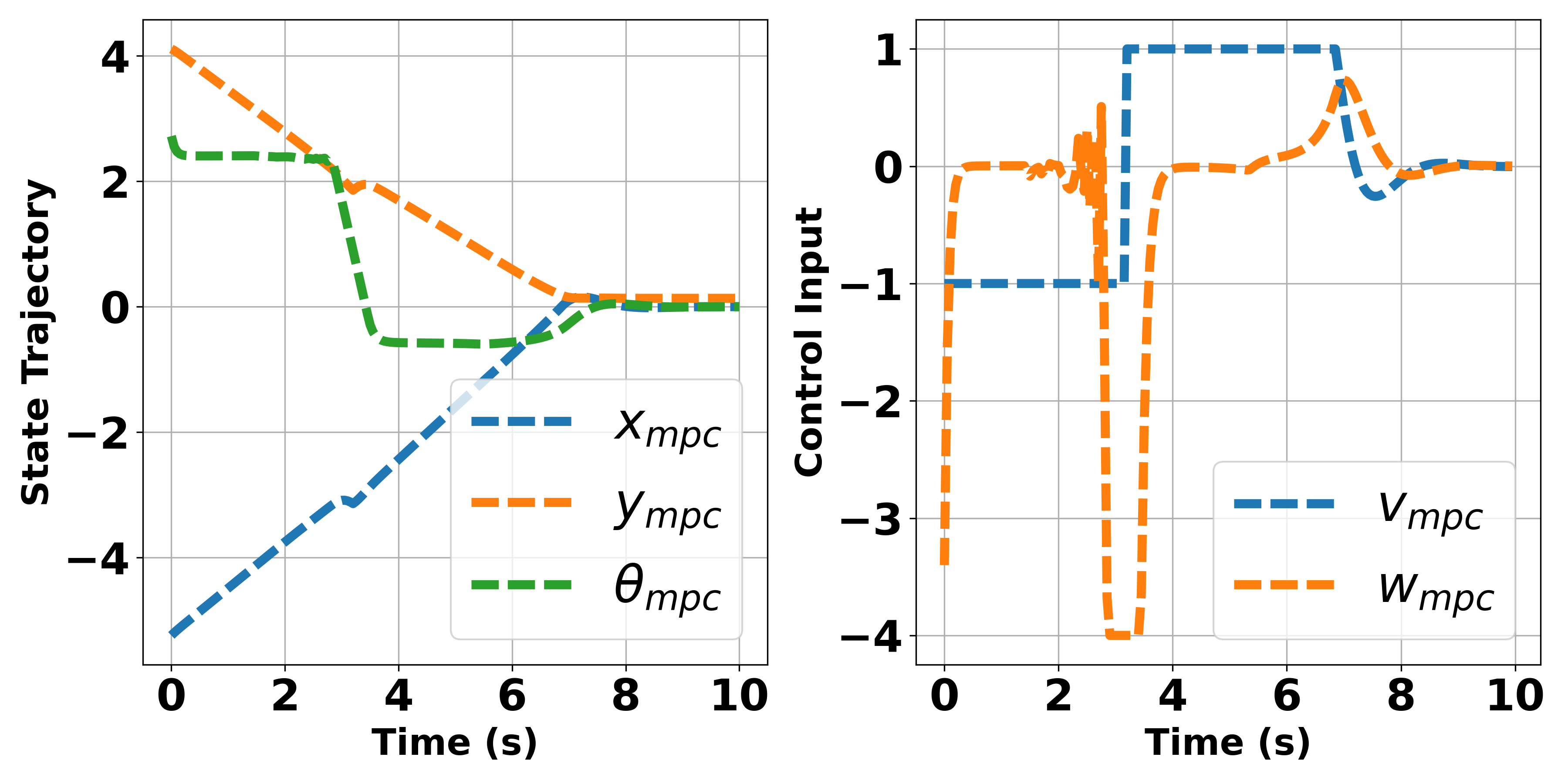}
   \caption{MPC state and control input trajectories}
   \label{fig:simulation2_mpc}
   \end{center}
\end{subfigure}
\begin{subfigure}[b]{1\columnwidth}
   \begin{center}
      \includegraphics[trim=0cm 0cm 0cm 0cm, width=1\textwidth]{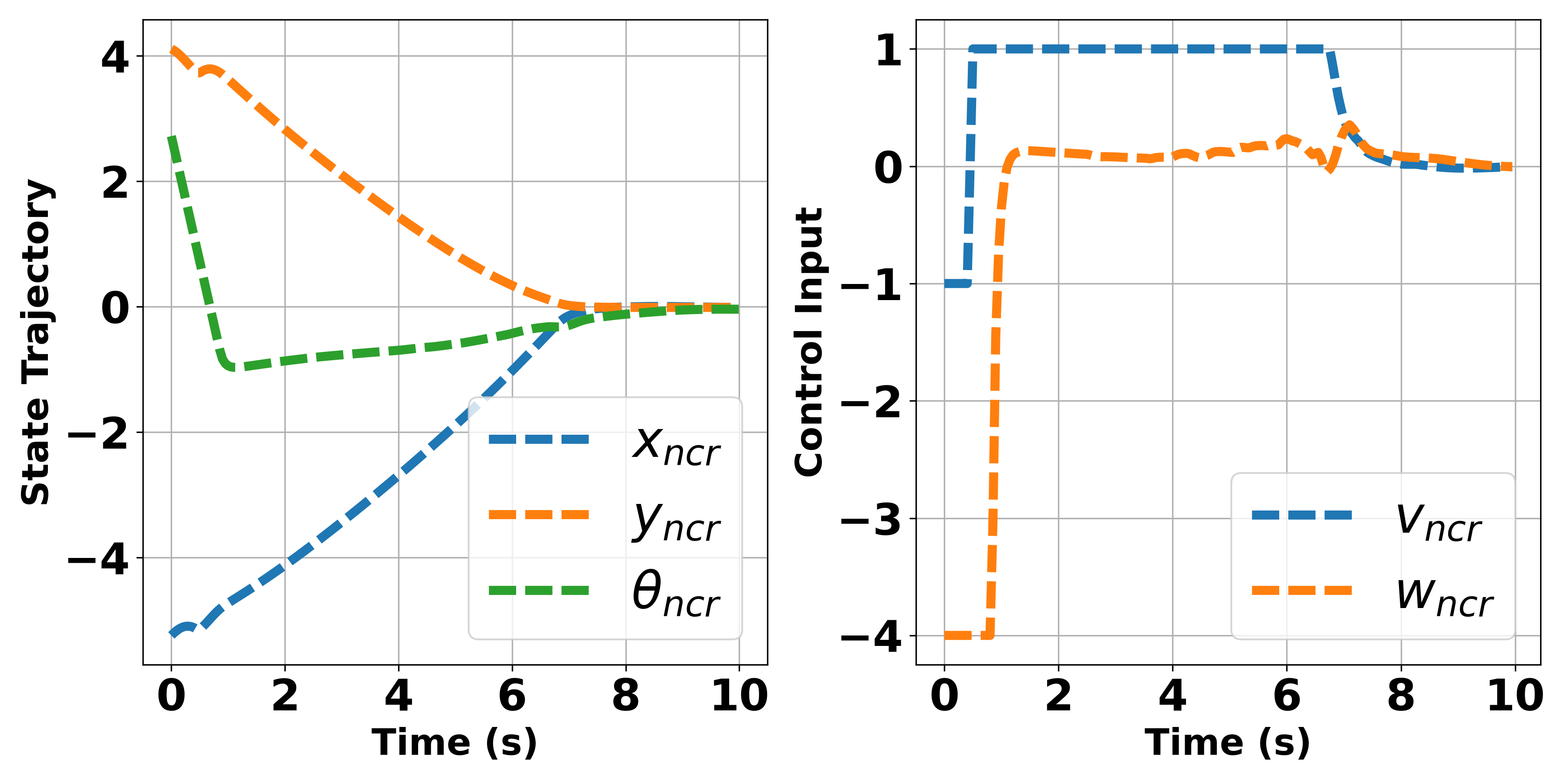}
\caption{NCR state and control input trajectories.}
   \label{fig:simulation2_ncr}
   \end{center}
\end{subfigure}
\caption[8pt]{Comparison of MPC and NCR solutions for unseen initial conditions and zero target reference.}\label{fig:simulation2}
\end{figure}

To validate the generalizability of NCR, we also tested it for the case when the initial condition is outside of the train data distribution. As shown in Fig. \ref{fig:simulation2}, $\mathbf{z}(0) = [-5.24, 4.11, 2.72]^\top$ is used as the initial state for both MPC and NCR, and results in the final state value $\mathbf{z}(t_f)_{ncr} = [-0.01, -0.01, 0.04]^\top$ for NCR with a speed of $1.6ms$ per step. In comparison, MPC has the final state $\mathbf{z}(t_f)_{mpc} = [0.00, 0.13, 0.00]^\top$ with a slower speed at $268.9ms$ per step, which is also worse than NCR in terms of the sum of the absolute value of the convergence error. As indicated in Fig. \ref{fig:simulation2}a, the input trajectories of the MPC also change more abruptly compared to the results from NCR.

\subsection{Constrained Control input with Unseen $\mathbf{z}(0)$ and Nonzero Reference}

\begin{figure}[h!]
\begin{subfigure}[b]{1\columnwidth}
   \begin{center}
   \includegraphics[trim=0cm 0.0cm 0.0cm 0.0cm,width=1\textwidth]{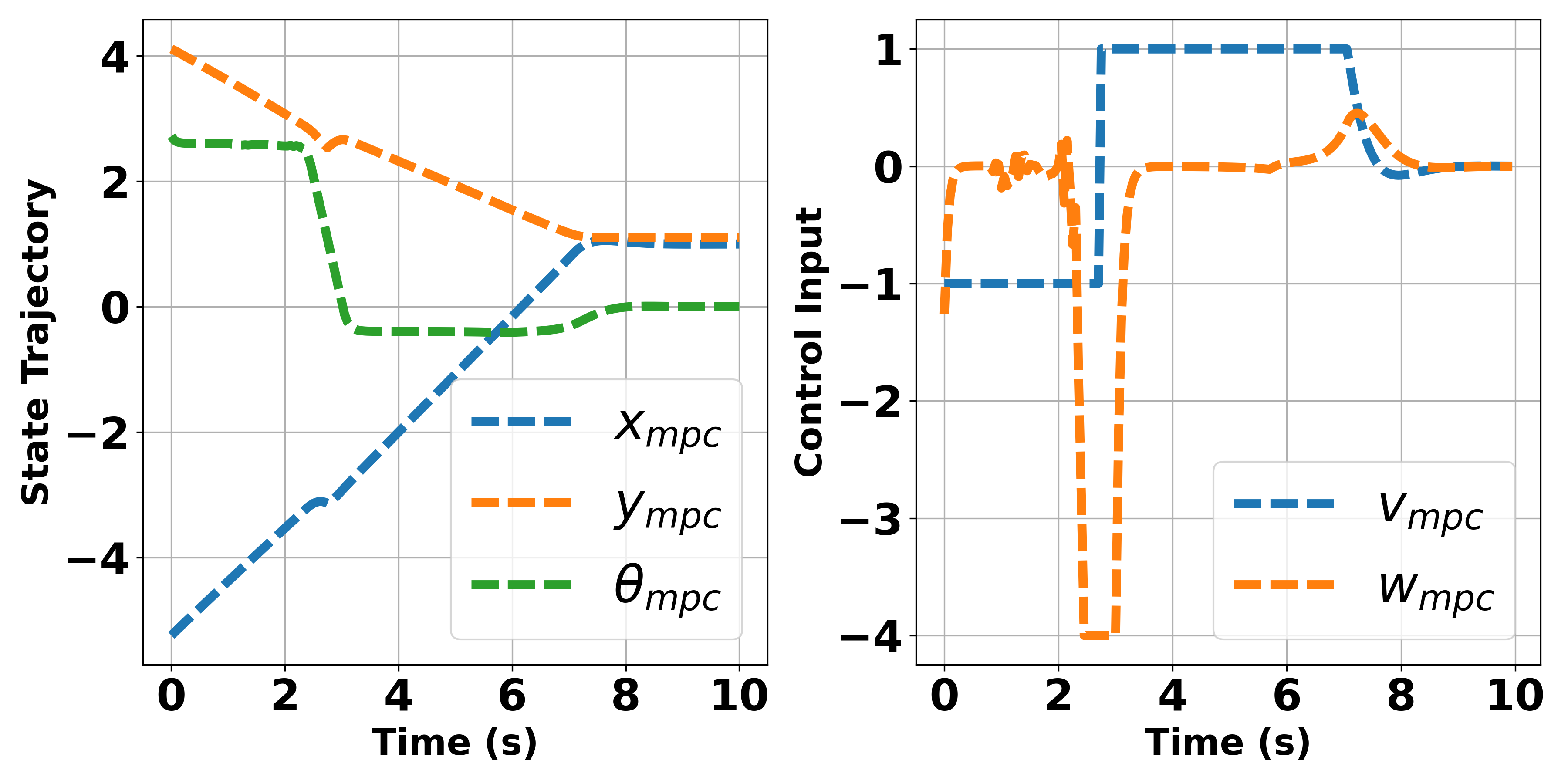}
   \caption{MPC state and control input trajectories.}
   \label{fig:simulation3_mpc}
   \end{center}
\end{subfigure}
\begin{subfigure}[b]{1\columnwidth}
   \begin{center}
      \includegraphics[trim=0cm 0cm 0cm 0cm, width=1\textwidth]{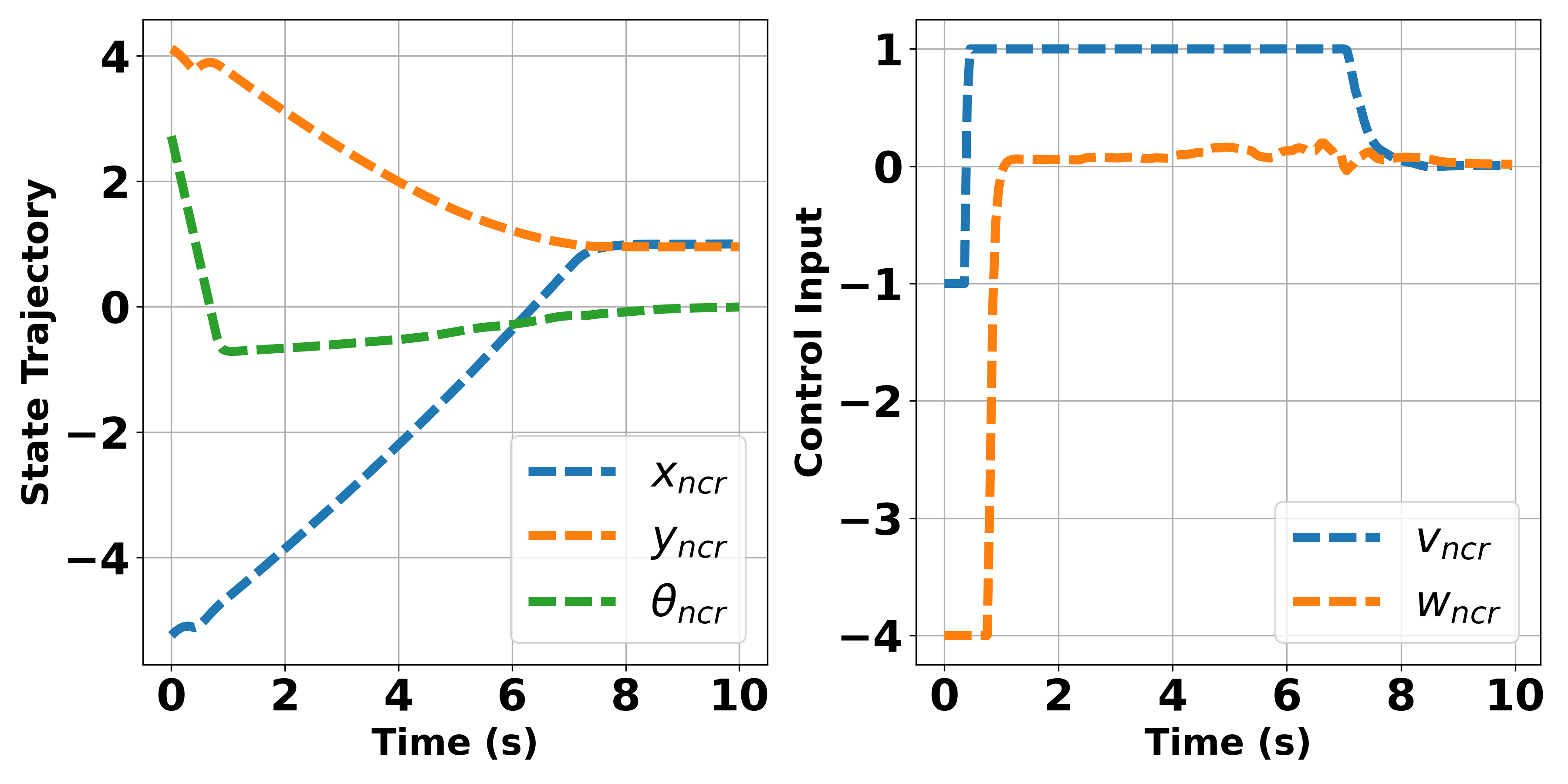}
\caption{NCR state and control input trajectories.}
   \label{fig:simulation3_ncr}
   \end{center}
\end{subfigure}
\caption[8pt]{Comparison of MPC and NCR solutions for unseen initial conditions and nonzero reference.}\label{fig:simulation3}
\end{figure}

We also use the same unseen initial state $\mathbf{z}(0) = [-5.24, 4.11, 2.72]^\top$ as in Subsection B, but the goal is to reach a nonzero reference state $\mathbf{z_{ref}} = [1.00, 1.00, 0.00]^\top$. {The input of the CoNN then becomes the error state ($\mathbf{z_k} - \mathbf{z_{ref}}$).} Our NCR is capable of reaching the target and ends up at the final state $\mathbf{z(t_f)_{ncr}} = [1.00, 0.95, 0.00]^\top$ with a consistent speed at $1.6ms$ per step. This further demonstrates its competency in dealing with an unseen initial state and reference target. For MPC, the final state is $\mathbf{z}(t_f)_{mpc} = [1.00, 1.11, 0.00]^\top$ and the speed is $275ms$ per step. The resulting state and control input trajectories of both NCR and MPC are shown in Fig. \ref{fig:simulation3}.

\subsection{Comparison with Model Predictive Control}

\begin{figure}[h!]
    \centering    \includegraphics[trim=0cm 0.0cm 0.0cm -0.8cm,width=0.5\textwidth]{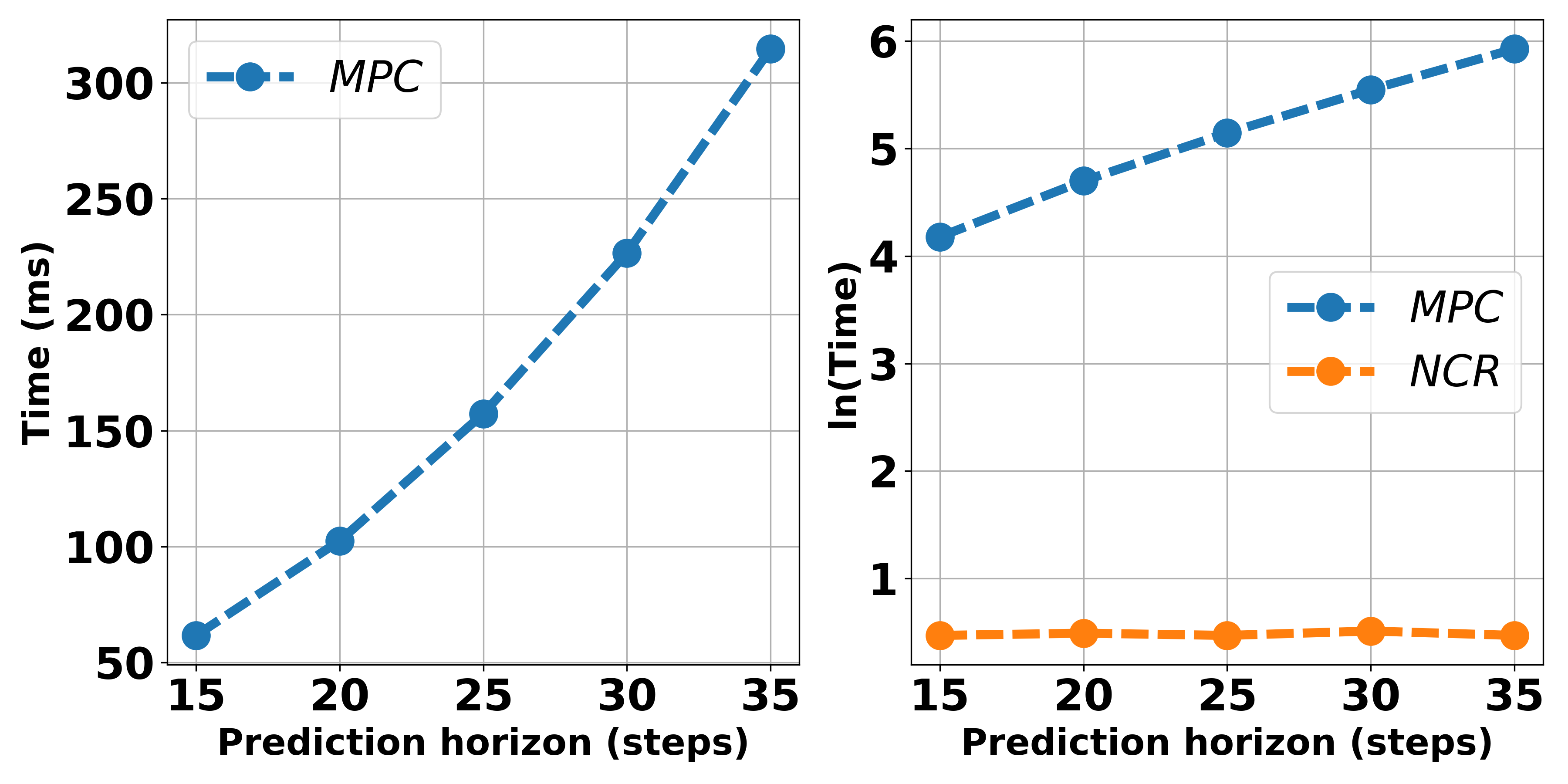}
    \caption{Computational time per simulation step with different prediction horizon for MPC (left) and natural log of time vs prediction horizon for both MPC and NCR (right).}
    \label{fig:time_per_step}
\end{figure}

To demonstrate the performance comparison of NCR and MPC quantitatively, the computational time for each simulation time step in cases of different prediction horizon is shown in Fig. \ref{fig:time_per_step}. Both MPC and NCR are tested using the same three previous examples, and the CoNN of NCR is trained with the corresponding prediction horizon off-line. For the sake of simpler notation, we denote the examples from previous subsections as case A, B, and C, respectively.

With regard to computational speed, we average the speed measured from three cases. As the prediction horizon increases, the computational speed of MPC increases almost exponentially, while the NCR remains almost constant, as shown in Fig. \ref{fig:time_per_step}. This makes sense because all NCR need is the CoNN inference time and solving a much simpler QP (based on the first predicted co-state vector), which has the same size for different prediction horizons. For the prediction horizon $n=30$, the computational speed of NCR is $100 \times$ faster than that of MPC. 

We also measured the trajectory smoothness using measure squared derivatives (MSD)\footnote{The mean squared derivative is computed by first calculating the numerical gradient at each point using the \textit{np.gradient} function, squaring each value, summing them, and dividing by the total number of points.}.
For both state and control input trajectories, which contain more than one trajectory, MSD is calculated by averaging the MSD value of each of their trajectory.
As indicated in Table \ref{tab:comparison}, the NCR is on par with MPC for state trajectories, but for both cases B and C, which have an initial state farther from the reference, the NCR input trajectory is significantly smoother. Note that MSD is calculated by taking average across the total simulation time steps, although the MPC state trajectory has a lower MSD value at case B and C, the MPC $\theta$ state trajectory has more abrupt changes in a shorter period of time around $3-4s$ and $7-8s$, which is less favorable than the result of NCR as shown in Fig. \ref{fig:simulation2} and Fig. \ref{fig:simulation3}. For each case, we sum the absolute value of convergence error of each state variable of the state vector, and the result is also shown in Table. \ref{tab:comparison}.

\begin{table}[h]
    \centering
    \renewcommand{\arraystretch}{1.2} % Increase row height for better readability
    \begin{tabular}{c c c c}
        \toprule
        \textbf{\makecell{Performance \\ Metric}} & 
        \textbf{Case A} & 
        \textbf{Case B} & 
        \textbf{Case C} \\
        \midrule
        \textbf{\makecell{MSD \\ (State)}} & \makecell{MPC: 0.19 \\ \textit{NCR: 0.16}}& \makecell{\textit{MPC: 0.62} \\ NCR: 0.71} & \makecell{\textit{MPC: 0.59} \\ NCR: 0.69} \\
        \midrule
        \textbf{\makecell{MSD \\ (Control Input)}} & \makecell{\textit{MPC: 2.21} \\ NCR: 2.40} & \makecell{MPC: 17.60 \\ \textit{NCR: 5.25}} & \makecell{MPC: 10.04 \\ \textit{NCR: 5.74}} \\
        \midrule
        \textbf{\makecell{Absolute \\ Convergence Error}} & \makecell{\textit{MPC: 0.14} \\ NCR: 0.16} & \makecell{MPC: 0.14 \\ \textit{NCR: 0.06}} & \makecell{MPC: 0.11 \\ \textit{NCR: 0.05}} \\
        \bottomrule
    \end{tabular}
    \caption{MPC and NCR performance comparison. \textit{Italicized} entries indicate better performance.}
    \label{tab:comparison}
    \vspace{-0.3cm} % Adjust space below the table if necessary
\end{table}

\section{Conclusion}\label{sec-conclusion}

Building upon the work of co-state neural network (CoNN), in this paper, we present a novel data-driven paradigm to train the CoNN in an unsupervised learning way for solving optimal control problems with input constraints, inspired by Deep Q-learning. The proposed \textit{neural co-state regulator} (NCR) not only greatly alleviates the online computational burden compared to MPC, but it can also be more generalizable to unseen scenarios. Future works can be made to have better design of loss function, in particular, the regularized co-state loss and better neural network architecture that may be potentially more suitable in different optimal control problems.


%%%%%%%%%%%%%%%%%%%%%%%%%%%%%%%%%%%%%%%%%%%%%%%%%%%%%%%%%%%%%%%%%%%%%%%%%%%%%%%%

%%%%%%%%%%%%%%%%%%%%%%%%%%%%%%%%%%%%%%%%%%%%%%%%%%%%%%%%%%%%%%%%%%%%%%%%%%%%%%%%

%%%%%%%%%%%%%%%%%%%%%%%%%%%%%%%%%%%%%%%%%%%%%%%%%%%%%%%%%%%%%%%%%%%%%%%%%%%%%%%%

\bibliographystyle{IEEEtran}
\bibliography{ref}

\begin{thebibliography}{10}
\providecommand{\url}[1]{#1}
\csname url@samestyle\endcsname
\providecommand{\newblock}{\relax}
\providecommand{\bibinfo}[2]{#2}
\providecommand{\BIBentrySTDinterwordspacing}{\spaceskip=0pt\relax}
\providecommand{\BIBentryALTinterwordstretchfactor}{4}
\providecommand{\BIBentryALTinterwordspacing}{\spaceskip=\fontdimen2\font plus
\BIBentryALTinterwordstretchfactor\fontdimen3\font minus \fontdimen4\font\relax}
\providecommand{\BIBforeignlanguage}[2]{{%
\expandafter\ifx\csname l@#1\endcsname\relax
\typeout{** WARNING: IEEEtran.bst: No hyphenation pattern has been}%
\typeout{** loaded for the language `#1'. Using the pattern for}%
\typeout{** the default language instead.}%
\else
\language=\csname l@#1\endcsname
\fi
#2}}
\providecommand{\BIBdecl}{\relax}
\BIBdecl

\bibitem{teo2021applied}
K.~L. Teo, B.~Li, C.~Yu, V.~Rehbock \emph{et~al.}, ``Applied and computational optimal control,'' \emph{Optimization and Its Applications}, 2021.

\bibitem{peaucelle2010complexity}
D.~Peaucelle and D.~Henrion, ``A survey of computational complexity results in systems and control,'' \emph{Automatica}, vol.~46, no.~7, pp. 1067--1084, 2010.

\bibitem{Betts2010}
J.~T. Betts, \emph{Practical Methods for Optimal Control and Estimation Using Nonlinear Programming}.\hskip 1em plus 0.5em minus 0.4em\relax Philadelphia, PA: SIAM, 2010.

\bibitem{Grune2011}
L.~Grüne and J.~Pannek, \emph{Nonlinear Model Predictive Control: Theory and Algorithms}, 1st~ed., ser. Communications and Control Engineering.\hskip 1em plus 0.5em minus 0.4em\relax New York: Springer, 2011.

\bibitem{schwenzer2021mpc}
\BIBentryALTinterwordspacing
M.~Schwenzer, M.~Ay, T.~Bergs, and D.~Abel, ``Review on model predictive control: An engineering perspective,'' \emph{Journal of Control, Automation and Electrical Systems}, vol.~32, no.~5, pp. 1214--1232, 2021.
\BIBentrySTDinterwordspacing

\bibitem{Nambisan2024Optimal}
\BIBentryALTinterwordspacing
P.~Nambisan and M.~Khanra, ``Optimal power-split of hybrid energy storage system using pontryagin’s minimum principle and deep reinforcement learning approach for electric vehicle application,'' \emph{Engineering Applications of Artificial Intelligence}, vol. 135, p. 108769, September 2024.
\BIBentrySTDinterwordspacing

\bibitem{kirk2004pontryagin}
D.~E. Kirk, \emph{Optimal Control Theory: An Introduction}, reprint edition~ed.\hskip 1em plus 0.5em minus 0.4em\relax Mineola, New York: Dover Publications, 2004, ch. 5.3, pp. 227--239, chapter 5.3: Pontryagin’s Minimum Principle and State Inequality Constraints.

\bibitem{rao2009survey}
A.~V. Rao, ``A survey of numerical methods for optimal control,'' \emph{Advances in the astronautical Sciences}, vol. 135, no.~1, pp. 497--528, 2009.

\bibitem{pagone2022penalty}
M.~Pagone, M.~Boggio, C.~Novara, A.~Proskurnikov, and G.~C. Calafiore, ``A penalty function approach to constrained pontryagin-based nonlinear model predictive control,'' in \emph{Proceedings of the 61st IEEE Conference on Decision and Control (CDC)}.\hskip 1em plus 0.5em minus 0.4em\relax IEEE, 2022, pp. 3705--3710.

\bibitem{bonalli2022sequential}
\BIBentryALTinterwordspacing
R.~Bonalli, T.~Lew, and M.~Pavone, ``Sequential convex programming for non-linear stochastic optimal control,'' \emph{ESAIM: Control, Optimisation and Calculus of Variations}, vol.~28, p.~64, 2022.
\BIBentrySTDinterwordspacing

\bibitem{pereira2021aggregated}
\BIBentryALTinterwordspacing
M.~de~Freitas Virgilio~Pereira, I.~V. Kolmanovsky, and C.~E.~S. Cesnik, ``Nonlinear model predictive control with aggregated constraints,'' \emph{Automatica}, vol. 132, p. 109746, 2021, brief paper.
\BIBentrySTDinterwordspacing

\bibitem{feng2024optimal}
\BIBentryALTinterwordspacing
M.~Feng, Z.~Chen, Y.~Huang, Y.~Liu, and J.~Yan, ``Optimal control operator perspective and a neural adaptive spectral method,'' \emph{arXiv preprint arXiv:2412.12469}, 2024, license: CC BY-NC-ND 4.0.
\BIBentrySTDinterwordspacing

\bibitem{sutton1998reinforcement}
R.~S. Sutton, A.~G. Barto \emph{et~al.}, \emph{Reinforcement learning: An introduction}.\hskip 1em plus 0.5em minus 0.4em\relax MIT press Cambridge, 1998, vol.~1, no.~1.

\bibitem{ddpg}
T.~P. Lillicrap, J.~J. Hunt, A.~Pritzel, N.~Heess, T.~Erez, Y.~Tassa, D.~Silver, and D.~Wierstra, ``Continuous control with deep reinforcement learning,'' \emph{arXiv preprint arXiv:1509.02971}, 2015.

\bibitem{ppo}
J.~Schulman, F.~Wolski, P.~Dhariwal, A.~Radford, and O.~Klimov, ``Proximal policy optimization algorithms,'' \emph{arXiv preprint arXiv:1707.06347}, 2017.

\bibitem{cobbe2019quantifying}
K.~Cobbe, O.~Klimov, C.~Hesse, T.~Kim, and J.~Schulman, ``Quantifying generalization in reinforcement learning,'' in \emph{International conference on machine learning}.\hskip 1em plus 0.5em minus 0.4em\relax PMLR, 2019, pp. 1282--1289.

\bibitem{underactuated}
\BIBentryALTinterwordspacing
R.~Tedrake, \emph{Underactuated Robotics}, 2023.
\BIBentrySTDinterwordspacing

\bibitem{chen2020imitation}
\BIBentryALTinterwordspacing
Y.~Chen, M.~Chen, and M.~Tomizuka, ``Imitation learning with neural network-based model predictive control,'' \emph{arXiv preprint arXiv:2001.02533}, 2020.
\BIBentrySTDinterwordspacing

\bibitem{bojarski2016endtoend}
M.~Bojarski, D.~D. Testa, D.~Dworakowski, B.~Firner, B.~Flepp, P.~Goyal, L.~D. Jackel, M.~Monfort, U.~Muller, J.~Zhang, X.~Zhang, J.~Zhao, and K.~Zieba, ``End to end learning for self-driving cars,'' \url{https://arxiv.org/abs/1604.07316}, 2016, nVIDIA Technical Report.

\bibitem{ross2011reduction}
S.~Ross, G.~Gordon, and D.~Bagnell, ``A reduction of imitation learning and structured prediction to no-regret online learning,'' in \emph{Proceedings of the fourteenth international conference on artificial intelligence and statistics}.\hskip 1em plus 0.5em minus 0.4em\relax JMLR Workshop and Conference Proceedings, 2011, pp. 627--635.

\bibitem{zhao2023finegrained}
T.~Z. Zhao, V.~Kumar, S.~Levine, and C.~Finn, ``{Learning Fine-Grained Bimanual Manipulation with Low-Cost Hardware},'' in \emph{Proceedings of Robotics: Science and Systems}, Daegu, Republic of Korea, July 2023.

\bibitem{effati2013optimal}
\BIBentryALTinterwordspacing
S.~Effati and M.~Pakdaman, ``Optimal control problem via neural networks,'' \emph{Neural Computing and Applications}, vol.~23, no. 7-8, pp. 2093--2100, 2013.
\BIBentrySTDinterwordspacing

\bibitem{pontryagin_nn-mathematics}
\BIBentryALTinterwordspacing
A.~D’Ambrosio, E.~Schiassi, F.~Curti, and R.~Furfaro, ``Pontryagin neural networks with functional interpolation for optimal intercept problems,'' \emph{Mathematics}, vol.~9, no.~9, 2021.
\BIBentrySTDinterwordspacing

\bibitem{zang2022machine}
Y.~Zang, J.~Long, X.~Zhang, W.~Hu, J.~Han \emph{et~al.}, ``A machine learning enhanced algorithm for the optimal landing problem,'' in \emph{Mathematical and Scientific Machine Learning}.\hskip 1em plus 0.5em minus 0.4em\relax PMLR, 2022, pp. 319--334.

\bibitem{lian2025co}
L.~Lian and U.~Inyang-Udoh, ``Co-state neural network for real-time nonlinear optimal control with input constraints,'' \emph{arXiv preprint arXiv:2503.00529}, 2025.

\bibitem{nonlinear-programming-book}
L.~T. Biegler, \emph{Nonlinear programming: concepts, algorithms, and applications to chemical processes}.\hskip 1em plus 0.5em minus 0.4em\relax SIAM, 2010.

\bibitem{numerical-tpbvp-book}
H.~B. Keller, \emph{Numerical solution of two point boundary value problems}.\hskip 1em plus 0.5em minus 0.4em\relax SIAM, 1976.

\bibitem{collocation-method-ocp}
S.~Oh and R.~Luus, ``Use of orthogonal collocation method in optimal control problems,'' \emph{International Journal of Control}, vol.~26, no.~5, pp. 657--673, 1977.

\bibitem{mehta2009qlearning}
P.~Mehta and S.~Meyn, ``Q-learning and pontryagin's minimum principle,'' in \emph{Proceedings of the 48th IEEE Conference on Decision and Control (CDC)}.\hskip 1em plus 0.5em minus 0.4em\relax IEEE, 2009, pp. 3598--3605.

\bibitem{deep-q-learning}
V.~Mnih, K.~Kavukcuoglu, D.~Silver, A.~A. Rusu, J.~Veness, M.~G. Bellemare, A.~Graves, M.~Riedmiller, A.~K. Fidjeland, G.~Ostrovski \emph{et~al.}, ``Human-level control through deep reinforcement learning,'' \emph{nature}, vol. 518, no. 7540, pp. 529--533, 2015.

\bibitem{rl_mpc_tnnls}
M.~Lin, Z.~Sun, Y.~Xia, and J.~Zhang, ``Reinforcement learning-based model predictive control for discrete-time systems,'' \emph{IEEE Transactions on Neural Networks and Learning Systems}, vol.~35, no.~3, pp. 3312--3324, 2024.

\bibitem{fmpc-iros}
M.~Greeff and A.~P. Schoellig, ``Flatness-based model predictive control for quadrotor trajectory tracking,'' in \emph{2018 IEEE/RSJ International Conference on Intelligent Robots and Systems (IROS)}, 2018, pp. 6740--6745.

\end{thebibliography}

\end{document}